\documentclass[twocolumn]{aa}
\usepackage{graphicx}
\usepackage{amsmath}
\usepackage{amsfonts}
\usepackage{amssymb}%
\begin{document}
   \title{Detectability of Cosmic Topology in Generalized Chaplygin Gas Models}

   \subtitle{}

   \author{B. Mota\inst{1}
          \and
          M. Makler
          \inst{1,2,3}
          \and
          M.J. Rebou\c{c}as\inst{1}}


   \institute{Centro Brasileiro de Pesquisas F\'{\i}sicas,
             Rua Dr.\ Xavier Sigaud, 150,
             22290-180 Rio de Janeiro -- RJ, Brazil\\
             \email{brunom@cbpf.br, reboucas@cbpf.br, martin@cbpf.br}
                         \and
             Universidade Federal do Rio de Janeiro,
             Instituto de F\'{\i}sica, C.P. 68528,
             21945-972 Rio de Janeiro -- RJ, Brazil
             \and
             Observat\'orio Nacional - MCT,
             Rua Gal. Jos\'e Cristino, 77
             20921-400, Rio de Janeiro -- RJ, Brazil
                           }

   \date{Received <date>; accepted <date>}

   \abstract{
   If the spatial section of the universe is multiply connected,
repeated images or patterns are expected to be detected
observationally. However, due to the finite distance to the last
scattering surface, such pattern repetitions could be unobservable.
This raises the question of whether a given cosmic topology is
detectable, depending on the values of the parameters of the
cosmological model. We study how detectability is affected by the
choice of the model itself for the matter-energy content of the
universe, focusing our attention on the generalized Chaplygin gas
(GCG) model for dark matter and dark energy unification, and
investigate how the detectability of cosmic topology depends on the
GCG parameters. We determine to what extent a number of topologies
are detectable for the current observational bounds on these
parameters. It emerges from our results that the choice of GCG as an
alternative to the $\Lambda$CDM  matter-energy content model has an
impact on the detectability of cosmic topology.

   }

   \maketitle
%

\section{Introduction}
Recent advances in observational cosmology have put stringent
constraints on the kind of universe we live in. Briefly, evidence
points to a cosmos which is homogeneous in large scales, with
spatial curvature close to zero, and where baryonic matter and
radiation contribute to only about 5\% of the it's total density.
The bulk of the universe's matter-energy content seems to be
composed of a pressureless component that is responsible for the
observed clustering of luminous matter (accounting for almost a
third of the universe's total density), as well as a negative
pressure component that drives the present phase of accelerated
expansion (resposible for the remaining two thirds of the total
density). Neither of these components can be directly observed and
they are usually referred to as dark matter (DM) and dark energy
(DE).

It has been proposed that both DE and DM may be described by a
single fluid, reducing to one the unknown components of the material
substratum of the universe. Among the candidates for such
\textit{unifying dark matter} are the Chaplygin gas, and
generalizations thereof (see for example Kamenshchik et al.
\cite{Kamenshchik}, Makler \cite{Makler}, and Bili\'{c} et al.
\cite{Bilic}). At high densities, such as those found at redshifts $z\gg1$ or
in galaxy clusters, this component behaves as DM, while at low
densities (such as the average density at $z\lesssim 1$) its
pressure becomes increasingly negative, explaining the current
accelerated expansion of the universe. Observational data constrain
the Chaplygin model's parameter space, but does not rule it out
(see, e.g., Makler et al. \cite{GCG2}, Reis et al. \cite{entropy},
Dev et al. \cite{Dev}, Zhu \cite{Zhu}, and references therein).

In a hitherto unrelated line of research, a great deal of work has
recently gone into studying the possibility that the universe may
possess compact spatial sections with a non-trivial topology (for
reviews, see for instance Lachi\`{e}ze-Rey et al.
\cite{CosmTopReviews}, Levin \cite{CosmTopReviews3}, and Rebou\c cas
\& Gomero \cite{RG2004}). Many methods of detecting such topology
have been proposed (see, e.g., Lehoucq et al. \cite{CitS}, Cornish
et al. \cite{PSH}, and Uzan et al. \cite{CCP}). The increasing
accuracy of observational cosmology now makes it possible to apply
these methods to existing observational data (see, e.g., Luminet et
al. \cite{Poincare}, Cornish et al. \cite{Cornishetal03}, de
Oliveira-Costa et al. \cite{CMB+NonGauss}, and Roukema et al.
\cite{Roukema}).

A direct way of determining the existence of a multiply connected
$3$-space section $M$ is by detecting repeated images of radiating
sources,
which implies that the separation between correlated pairs must be
smaller than twice the radius of the observable universe, $d_{hor}$.
Note however that this distance between images, being the length of
the associated closed geodesic, is fixed for each nonflat topology
in units of the curvature radius $a_0$. As a consequence, many
possible topologies may be undetectable by pattern repetition, given
the nearly flatness of the universe favored by current observations
(see, e.g., Tegmark et al. \cite{SDSSWMAP}). Thus, the ratio
$\chi_{hor}=d_{hor}/a_0$, which depends on the cosmological model
and its associated parameters, is crucial to determine the
detectability of cosmic topology.

The extent to which a nontrivial topology may or may not be detected
for the current bounds on the cosmological parameters is studied in
Gomero et al. (\cite{Topdetect}, \cite{Topdetect2}), Mota et al.
(\cite{Mota03}), Weeks (\cite{NEWweeks}), and Weeks et al.
(\cite{NEWweeks2}). These studies have concentrated on cases where
the matter-energy content is modeled within the $\Lambda$CDM
framework, investigating the dependence of the detectability of the
topology on the cosmological density parameters $\Omega_{m}$ and
$\Omega_{\Lambda}$. Here we shall address a different question: to
what extent is the detectability of the topology modified when
different background cosmological models are considered? To this
end, we shall focus our attention on low curvature nonflat universes
whose dark-matter-energy content is dominated by the generalized
Chaplygin Gas (GCG). We determine which topologies from a large
family of spherical manifolds and some small-volume hyperbolic
manifolds are either potentially detectable or undetectable, taking
into account the current observational bounds on the GCG model. We
also study the dependence of detectability on the model parameters,
and in particular on the steepness of the equation of state,
$\alpha$, and show that detectability becomes more likely as
$\alpha$ decreases.

This paper is organized as follows. In the next section we discuss
some fundamental results on the detectability of cosmic topology
and provide a short review of GCG-dominated cosmological models.
These two strands are brought together in section (\ref{detect}),
which deals with our analytical and numerical results regarding
the detectability of cosmic topology in the GCG case. Finally, we
sum up our results and indicate points for future research in
section (\ref{conclusion}).

\section{Preliminaries\label{prelim}}
Most cosmologists agree that the universe is close to being
spatially  homogeneous and isotropic at large scales. In a general
relativistic context, such a space-time is described by a
$4$-manifold $\mathcal{M}$, which is decomposed into $\mathcal{M}
= \mathbb{R} \times M$, and is endowed with a Robertson-Walker
metric
\begin{equation}
\label{RWmetric} ds^2 = -dt^2 + a^2 (t) \left [ d \chi^2 +
f^2(\chi) (d\theta^2 + \sin^2 \theta  d\phi^2) \right ] \;,
\end{equation}
where $a(t)$ is the scale factor and $f(\chi)=\chi\,$, $\sin\chi$,
or $\sinh\chi$, depending on the sign of the curvature (respectively
$k=0,1,$ or $-1$). For each of these three cases the spatial section
$M$ is usually taken to be a simply-connected manifold, either
euclidian $\mathbb{E}^{3}$ ($k=0$), spherical $\mathbb{S}^{3}$
($k=1$), or hyperbolic $\mathbb{H}^{3}$ ($k=-1$). However, since
geometry does not dictate topology, the $3$-space $M$ may as well be
any one of a number of (multiply connected) quotient manifolds $M =
\widetilde{M}/\Gamma$, where $\Gamma$ is a discrete and fixed
point-free group of isometries of the covering space
$\widetilde{M}=(\mathbb{E}^{3},\mathbb{S}^{3}, \mathbb{H}^{3})$. The
action of $\Gamma$ tessellates $\widetilde{M}$ into identical
domains which are copies of what is known as the fundamental
polyhedron. Hence, a point $x$ (or source) within the fundamental
polyhedron can be connected to every other point in its orbit in the
covering space $gx|_{g\in\Gamma}$ (i.e., its images), by a
space-like geodesic, which is closed by the associated isometry. The
immediate observational consequence of this is that the sky may
(potentially) show multiple images of radiating sources. The closed
geodesics are the projection onto $M$ of the (light-like) geodesics
followed by photons emitted by the source. By observing repeated
images or patterns, we are directly detecting isometries in
$\Gamma$, with which we might reconstruct $M$ and thus determine the
topology of the universe.

A closed geodesic that passes through $x$ in a multiply connected
manifold $M$ is a segment of a geodesic in the covering space
$\widetilde{M}$ that joins $x$ to one of its images. Since any
such pair of points is by definition related by an isometry $g \in
\Gamma$, the length of the closed geodesic associated to any fixed
isometry $g$ passing through $x$ is given by the corresponding
distance function $\delta_g(x) \equiv  d(x,gx)\,$. The injectivity
radius at $x$ is then defined as
\begin{equation}
\label{rinjx} r_{inj}(x) \equiv \frac{1}{2} \min_{g \in
\Gamma,\,g\neq I}\, \{\,\delta_g(x)\,\} \; .
\end{equation}
where $I$ is the identity of $\Gamma$. From (\ref{rinjx}) clearly
the injectivity radius is the radius of the smallest sphere
centered at $x$ and `inscribable' in the fundamental polyhedron.
Thus, any sphere with radius $r < r_{inj}(x)$ and centered at $x$
lies inside a fundamental polyhedron of $M$ based on $x$.

One can also define the global injectivity radius (simply denoted by
$r_{inj}$), which is the radius of the smallest ``inscribable''
sphere centered at any point in $M$, or the minimum value of
$r_{inj}(x)$
\begin{equation}
\label{rinj} r_{inj} \equiv  \min_{x \in M} \{\,r_{inj}(x)\,\} \;
.
\end{equation}

Due to the finite age of the universe, it is generally not
possible to observe the entire covering space. For any given
observational survey up to a maximum redshift $z_{obs}$, the
survey depth $\chi(z_{obs},\Omega_{i0})$ in a nonflat universe can
be expressed, in units of the curvature radius\footnote{Here $H_0$
is the Hubble constant and
$\Omega_{0}=\rho_{0}/\rho_{\mathrm{crit}}$ is the total density
parameter, with $\rho_{\mathrm{crit}}=3H_{0}^{2}/8\pi G$.}
$a_0=H_0/\sqrt{|1-\Omega_0|}$, as a function of the (present time)
density parameters $\Omega_{i0}=\rho_{i0}/\rho_{\mathrm{crit}}$,
as
\begin{equation}
\chi(z_{obs},\Omega_{i0})=\frac{d_{hor}}{a_0}=\sqrt{|1-\Omega_{0}|}
\int_{0}^{z_{obs}} \frac{H_0}{H(z,\Omega_{i0})} dz\;,\label{rd}
\end{equation}
where $H(z,\Omega_{i0})$ is the Hubble parameter at redshift $z$.

Now, for a given $\chi(z_{obs},\Omega_{i0})$ (or simply
$\chi_{obs}$) a topology is undetectable by an observer at a point
$x$ if $\chi_{obs} < r_{inj}(x)$.\footnote{In this work we shall
express $r_{inj}$ in units of $a_0$.} %
In this case there are no multiple
images (or pattern repetitions) in the survey. On the other hand,
when $\chi_{obs}> r_{inj}(x)$, then the topology is detectable in
principle for an observer at $x$. However, since we do not know
\textit{a priori} our position in the universe, it is more suitable
for our purposes here to use a criterium valid at any $x$,
stated in terms of the global injectiviy radius $r_{inj}$, which is
a lower bound for $r_{inj}(x)$. The undetectability
condition then is that the topology of $M$ is undetectable by any
observer if $r_{inj} > \chi_{obs}$. Reciprocally, the condition
$\chi_{obs} > r_{inj}$ ensures detectability in principle for at
least some observers. In globally homogeneous manifolds, the
distance function $d(x,gx)$ for any given isometry $g$ is constant,
which means $r_{inj}(x)\equiv r_{inj}$.
Therefore, if the topology is potentially detectable (or conversely,
undetectable) by an observer at $x$, it is likewise potentially
detectable (conversely undetectable) by any other observer at any
other point in the $3$-space $M$.

We shall focus exclusively on the detectability of spherical and
hyperbolic manifolds. These manifolds are rigid,
which implies that geometrical quantities are topological invariants
and therefore the
lengths of the closed geodesics are fixed in units of the
curvature radius $a_0$. We can therefore compare the injectivity
radius $r_{inj}$ of each manifold with the survey depth
$\chi_{obs}$ as a function of the matter-energy content model
parameters.

The spherical $3$-manifolds are of the form $M =
\mathbb{S}^3/\Gamma$, where $\Gamma$ is a finite subgroup of
$SO(4)$ acting freely on the $3$-sphere. These manifolds were
originally classified  by Threlfall and Seifert
(\cite{ThrelfallSeifert}), and are also discussed by Wolf
(\cite{Wolf}) (for a description in the context of cosmic topology
see Ellis \cite{Ellis71}). This classification consists
essentially in the enumeration of all finite groups $\Gamma
\subset SO(4)$, and then grouping all ensuing manifolds in
classes. In a recent paper, Gausmann et al. (\cite{GLLUW}) recast
the classification in terms of single action, double action, and
linked action manifolds. In table~\ref{SingleAction} we list the
orientable single action manifolds together with the labels often used to
refer to them, as well as the order of the covering groups
$\Gamma$ and the corresponding injectivity radii. All single
action manifolds are globally homogeneous, and thus the same
detectability condition holds for all observers in the manifold.
%
\begin{table}[!htb]
\caption{Single action spherical manifolds along with the order of
the covering groups and the injectivity radii.}
\label{SingleAction}
\centering
\begin{tabular}{ l c c c }
\hline\hline
 \ Name \&  Symbol \ &  \ Order of $\Gamma$ \ & \
Injectivity Radius \  \\ \hline Cyclic              $Z_n$ & $n$ &
$\pi/n$           \\ Binary dihedral     $D^*_m$ & $4m$ & $\pi / 2m
$      \\ Binary tetrahedral  $T^*$   & 24   & $ \pi/6$
\\ Binary octahedral   $O^*$   & 48   & $\pi/8$          \\ Binary
icosahedral  $I^*$   & 120  & $\pi/10$         \\ \hline
\end{tabular}
\end{table}

Closed orientable hyperbolic $3$-manifolds can be constructed and
studied with the publicly available SnapPea software package %
(Weeks \cite{SnapPea}). Each compact hyperbolic manifold is
constructed from a non-compact cusped manifold through a so-called
Dehn surgery, which is a formal procedure identified by two
coprime integers called winding numbers $(n_1,n_2)$. SnapPea names
these manifolds according to the original cusped manifold and the
winding numbers. So, for example, the smallest volume hyperbolic
manifold known (the Weeks' manifold) is labeled as m$003(-3,1)$,
where m$003$ is the seed cusped manifold and $(-3,1)$ are the
winding numbers. Hodgson and Weeks (\cite{HodgsonWeeks}) have
compiled a census containing $11031$ orientable closed hyperbolic
3-manifolds ordered by increasing volume. In
table~\ref{10HW-Census} we present the seven smallest manifolds
from this census, with their respective volumes and injectivity
radii $r_{inj}$.
%
\begin{table}[!htb]
\caption{First seven manifolds in the Hodgson-Weeks census of
closed hyperbolic manifolds, with corresponding volumes and
injectivity radii.} \label{10HW-Census}
\centering
\begin{tabular}{l c c c }
\hline\hline \ \ \ Manifold \ \ \ & \ \ \ Volume \ \ \ & \ \ \
Injectivity Radius  \\ \hline
\ \ m003(-3,1) &  0.943  & 0.292  \\ 
\ \ m003(-2,3) &  0.981  & 0.289  \\ 
\ \  m007(3,1) &  1.015  & 0.416  \\ 
\ \ m003(-4,3) &  1.264  & 0.287  \\ 
\ \  m004(6,1) &  1.284  & 0.240  \\ 
\ \  m004(1,2) &  1.398  & 0.183  \\ 
\ \  m009(4,1) &  1.414  & 0.397  \\ \hline
\hline
\end{tabular}
\end{table}

It is clear from the equation (\ref{rd}) that, since the integral
is finite, as $\Omega_{0}\rightarrow1$ more and more nonflat
topologies become undetectable. To gain a more quantitative
understanding of the issue of detectability, however, one must
first specify the equations of state of the component densities.
Therefore, the choice of model for the matter-energy content of
the universe has, as we shall discuss below, important
consequences for the possible detection of cosmic topology. In the
present work we study the detectability of cosmic topology in a
universe dominated by the GCG, which we now briefly review.

As mentioned in the introduction, the dynamics of the universe
seems to be dominated by two components which are not directly
observable: a pressureless DM component and a DE component
with negative pressure. DM is detected by its local clustering,
through dynamical measurements, such as galaxy rotation curves,
velocity dispersion and $X$-ray emission by gas in clusters, and
also by gravitational lensing. The observed abundance of light
elements together with primordial nucleosynthesis calculations
show that baryons can account for only about 15\% of this
clustered component (see, e.g., Bertone et al. \cite{DMreviews}).
On the other hand, the presence of DE is evidenced by its effects
on very large scales. It seems to power the accelerated expansion
discovered in type Ia supernovae (SNIa) data (see, e.g.,
Perlmutter et al. \cite{PerlmutterSNIa}, Riess et al.
\cite{Riess}, and Tonry et al. \cite{tonry03}) and to
significantly contribute to the global curvature --- providing
around two thirds of the density needed to explain the near
flatness inferred from the cosmic microwave background radiation
(CMBR) data.

Since the evidence for DM and DE involves observations at
different scales and epochs, one may wonder if they could be
distinct manifestations of the same component, dubbed unifying
dark matter, or UDM (see, e.g., Makler et al. \cite{GCG1}). We
will assume that the UDM may be modeled by a perfect fluid
energy-momentum tensor whose only independent components are $p$
and $\rho$. The UDM equation of state must be such as to allow for
a decelerating and non-relativistic phase in the past, followed by
the present accelerated phase. Recall that the acceleration is
given by $\ddot{a}=-4\pi G (\rho+3p)/3$. Thus, for the
decelerating and accelerating phases to happen sequentially the
UDM must be nearly pressureless at higher densities, with
$|p|\ll\rho$, and also exert a large ($p\sim-\rho$) negative
pressure at lower densities. A simple example of an equation of
state satisfying this condition is that of the GCG (see Makler
\cite{Makler}, Bili\'{c} et al. \cite{Bilic}, and Bento et al.
\cite{BentoGCG}), and is given by an inverse power law
\begin{equation}
p_{\mathrm{Ch}}=-\frac{M^{4(\alpha+1)  }}{\rho_{\mathrm{Ch}%
}^{\alpha}}\;,\label{eostoy}%
\end{equation}
where $M$ is positive and has dimension of mass and $\alpha$ is an
adimensional real number.

Consider now the background geometry of a universe dominated by
the GCG. In this case, the energy conservation equation can be
easily solved, giving the energy density of the GCG in terms of
the scale factor,
\begin{equation}
\rho_{\mathrm{Ch}}=\rho_{\mathrm{Ch}0}\left[(1-A)\left(  \frac{a_{0}}%
{a}\right)  ^{3(\alpha+1)}+A\right]  ^{1/(\alpha+1)},\label{rhoa}%
\end{equation}
where $A=(M^{4}/\rho_{\mathrm{Ch}0})^{(\alpha+1)}$. We must impose
$\alpha>-1$ for the accelerated epoch to occur after the
decelerated phase. We also require $0<A<1$ so that
$\rho_{\mathrm{Ch}}$ is well defined at all times. Under these
circumnstances, at earlier times when $a/a_{0}\ll1$, we have
$\rho_{\mathrm{Ch}0}\propto a^{-3}$ and the fluid behaves as DM.
For later times, $a/a_{0}\gg1$, and we have
$p_{\mathrm{Ch}0}=-\rho_{\mathrm{Ch}0}=-M^{4}=const.$ as in the
cosmological constant case. Thus, this equation of state
interpolates between DM and DE while the average energy density of
the universe changes, as expected from the previous discussion. It
includes as special cases the standard Chaplygin gas for
$\alpha=1$, and the $\Lambda$CDM model for $\alpha=0$.

We proceed now to study the interplay between the detectability of
cosmic topology and the GCG parameters $A$ and $\alpha$.

\section{Detectability of topology in GCG cosmology\label{detect}}

The key point for a systematic and quantitative study of the
detectability of the topology of the (nonflat) spatial section $M$
is the comparison between the injectivity radius of each manifold
in units of the curvature radius $r_{inj}^{M}$ (a topological
invariant) with the survey depth $\chi_{obs}$. As was pointed out
in the previous section, the manifold's topology is detectable in
principle for a survey of depth $\chi_{obs}$ if the matter-energy
content parameters are such that $\chi_{obs}\geq r_{inj}^{M}$.
Likewise, if $\chi_{obs}<r_{inj}^{M}$ then the topology is
undetectable by any pattern repetition method. Most of the work
done so far focused on the $\Lambda$CDM model (see, e.g., Mota et
al. \cite{Mota03}), but here we extend this approach to GCG
models.

For the $\Lambda$CDM case (equivalent to $\alpha=0$ in the GCG
model) the detectability conditions can be restated in terms of
contour curves in the $\Omega_{m} -\, \Omega_{\Lambda}$ parametric
plane. For each manifold $M$ and a given fixed redshift $z_{obs}$ we
can define the contour curve
$\chi(\Omega_{m},\Omega_{\Lambda},z_{obs})=r_{inj}^{M}$. This curve
lies in either the positive or negative curvature semiplanes
($\Omega_{0}<1$ and $\Omega_{0}>1$ respectively), depending on
whether the manifold is spherical or hyperbolic. The contour curve
divides its semiplane in a region where the topology is undetectable
($\chi_{obs}<r_{inj}^{M}$), and a region where the topology is
detectable in principle ($\chi_{obs}\geq r_{inj}^{M}$). Therefore,
given this curve, it is possible to determine the (un)detectability
of any given nonflat manifold for a range of density parameters.

In a previous work (Mota et al. \cite{Mota03}) it was shown that
this contour curve can be approximated in two complementary ways by
the tangent and secant lines, which respectively overestimate and
underestimate detectability. It was further shown that the numerical
results from both methods are in good agreement with each other for
the bounds on density values obtained by Spergel et al.
(\cite{WMAP-Spergel-et-al}), which means they are good
approximations for the contour curve. Finally, it was also shown
that in the limit $z \rightarrow \infty$, the secant line can be
obtained analytically. This last result is of particular interest,
because it allows to study the detectability of topology not only
for individual manifolds, but also for whole classes of manifolds.

While in the $\Lambda$CDM case the detectability was determined by
two parameters, in the case of the GCG three parameters must be
taken into account,
$\Omega_{\mathrm{Ch}0}=\rho_{\mathrm{Ch}0}/\rho_{crit}$, $\,A$,
and $\alpha$ (with the baryonic density $\Omega_{b0}$ kept fixed).
To make the analysis simpler and comparisons easier, we introduce
the new variables
\begin{eqnarray}
\label{newv}%
\bar{\Omega}_{\Lambda0} & = &\Omega_{\mathrm{Ch}0}\,A\;,\nonumber\\
\bar{\Omega}_{m0} & = &\Omega_{\mathrm{Ch}0}\,(1-A)\;,
\end{eqnarray}
such that
$\Omega_{\mathrm{Ch}0}=\bar{\Omega}_{m0}+\bar{\Omega}_{\Lambda 0}$.
Notice that with this definition, for $\alpha=0$,
$\bar{\Omega}_{m0}$ corresponds to a matter density parameter and
$\bar{\Omega}_{\Lambda0}$ corresponds to a cosmological constant
parameter (see eq.~\ref{chiGCG2} bellow).

With the definitions (\ref{newv}) the redshift-distance relation
(\ref{rd}) becomes
\begin{eqnarray}
\chi_{obs}  &&=   \sqrt{|1-\Omega_0|}\times \nonumber\\
\int_{0}^{z}dx&& \left\{\Omega_{\mathrm{Ch}} + \Omega_{b0}(1+x)^{3}+%
(1-\Omega_0)(1+x)^{2}\right\} ^{-\frac{1}{2}}, \label{chiGCG2}
\end{eqnarray}
where
\begin{equation}
\Omega_{\mathrm{Ch}}=\Omega_{\mathrm{Ch}0}\left[  \frac{\bar{\Omega}_{\Lambda0}%
}{\Omega_{\mathrm{Ch}0}}+\frac{\bar{\Omega}_{m0}}{\Omega_{\mathrm{Ch}0}}(1+x)^{3(1+\alpha
)}\right]^{\frac{1}{1+\alpha}} \label{OmCh}
\end{equation}
and
$\Omega_0=\Omega_{\mathrm{Ch}0}+\Omega_{b0}$.

In the analytical computations that follow, we shall omit the
baryon fraction. The role of $\Omega_{b0}$ will be discussed later
in conjunction with the numerical computations. Thus, for a fixed
$z_{obs}$ the only free parameters left are $\alpha$,
$\bar{\Omega}_{m0}$, and $\bar{\Omega}_{\Lambda0}$. For each value
of $\alpha$ we can then define a parametric plane
$\bar{\Omega}_{m0}-\bar{\Omega}_{\Lambda0}$, which is very similar
to the $\Omega_{m0}-\Omega_{\Lambda0}$ plane discussed above.
Unfortunately, it is not possible to analytically integrate equation
(\ref{chiGCG2}) for a generic value of $\alpha$. Instead, we first
show that $\chi_{obs}$ is a monotonically decreasing function of
$\alpha$ for $\alpha\in(-1,\infty)$, and then obtain analytical
expressions for the contour curves for the limiting cases
$\alpha=-1$ and $\alpha\rightarrow\infty$, with
$z_{obs}\rightarrow\infty$. This sets upper and lower bounds on
the values $\chi_{obs}$ can take in equation (\ref{chiGCG2}), and
guarantees that $\chi_{obs}$ will interpolate monotonically
between these bounds for the intermediate values of $\alpha$.

Let us first show the monotonicity of $\chi_{obs}$ with respect to
$\alpha$. Consider the term $\Omega_{\mathrm{Ch}}$ given by
expression (\ref{OmCh}), with $x\geq0$ and
$\bar{\Omega}_{\Lambda0}\neq0$. Defining $u=(1+x)^{3(1+\alpha)} \ge
1$, a straightforward calculation shows that
\begin{equation}
\frac{\partial^{2}\;\Omega_{\mathrm{Ch}}}{\partial\;u\;\partial\;\alpha}=\frac
{\bar{\Omega}_{m0}}{\Omega_{\mathrm{Ch}0}}\log\left[
\frac{u}{\frac{\bar{\Omega
}_{\Lambda}}{\Omega_{\mathrm{Ch}0}}+\frac{\bar{\Omega}_{m}}{\Omega_{\mathrm{Ch}0}}u}\right]
\geq0\;.
\end{equation}
The derivative $\partial\;\Omega_{\mathrm{Ch}}/\partial\;\alpha$ is
therefore an increasing function of $u$. But for $u=1$, clearly
$\partial\;\Omega _{\mathrm{Ch}}/\partial\;\alpha=0$. Hence,
$\partial\;\Omega_{\mathrm{Ch}}/\partial \;\alpha>0$ for any $u>1$,
and $\Omega_{\mathrm{Ch}} $ is an increasing function of $\alpha.$\
Thus, from (\ref{chiGCG2}) we have that $\chi_{obs}$ is a decreasing
function of $\alpha$. As a result, the detectability of a given
topology then becomes less likely as $\alpha$ increases, for fixed
$\bar{\Omega}_m$ and $\bar{\Omega_{\Lambda}}$.

We shall now obtain analytical expressions for the contour curves in
the extreme cases $\alpha=-1$ and $\alpha=\infty$ (with
$\Omega_{b0}=0$). In both cases we take the
$z_{obs}\rightarrow\infty$ limit, which is an upper bound to the
size of the observable universe and does not give rise to
significant numerical differences from $z_{obs}=1089$ (which
corresponds to the last scattering surface).

When we take the $\alpha\rightarrow\infty$ limit in equation
(\ref{chiGCG2}), it is clear that $\chi_{obs}$ becomes a function of
$\Omega_{\mathrm{Ch}0}$ only. We then solve the equation
$\chi_{obs}(\Omega_{\mathrm{Ch}0})=r_{inj}$ for
$\Omega_{\mathrm{Ch}0}$ to obtain

\begin{eqnarray}
\label{condinf}%
\bar{\Omega}_{m0}+\bar{\Omega}_{\Lambda 0} & = & \mathrm{sech}^{2}\left(%
\frac{r_{inj}}{2}\right)
\;,\quad\;\;\mathrm{for}\quad\Omega_{\mathrm{Ch}0}<1\;,\nonumber\\
\bar{\Omega}_{m0}+\bar{\Omega}_{\Lambda 0} & = &
\mathrm{sec}^{2}\left( \frac{r_{inj}}{2}\right) \;,\quad\quad
\mathrm{for}\quad\Omega_{\mathrm{Ch}0}>1\;.\\&   &
\quad\quad\quad\quad\quad\quad\quad\quad\quad\quad\quad\quad
(\alpha\rightarrow\infty)\nonumber
\end{eqnarray}
The introduction of a small non-zero value of $\Omega_{b0}$ does
not change this result significantly. The undetectability
condition derived from the contour curves (\ref{condinf}) can be stated as:
\emph{for $\alpha\rightarrow\infty$ the manifold $M$ with
injectivity radius $r_{inj}^{M}$ is undetectable if either}
\begin{eqnarray}
\label{condinf2}%
\bar{\Omega}_{m0}+\bar{\Omega}_{\Lambda 0} & > & \mathrm{sech}^{2}\left(%
\frac{r_{inj}^{M}}{2}\right)
\;\;\;\mathrm{and}\quad\Omega_{\mathrm{Ch}0}<1\;,\nonumber\\
\mathrm{or}\\
\bar{\Omega}_{m0}+\bar{\Omega}_{\Lambda 0} & <
&\mathrm{sec}^{2}\left( \frac{r_{inj}^{M}}{2}\right) \;\quad
\mathrm{and}\quad\Omega_{\mathrm{Ch}0}>1\;.\nonumber
\end{eqnarray}

The calculation in the limit $\alpha\rightarrow-1$ is somewhat
more involved.
Let $\epsilon=\alpha-1$. We have, up to first order in $\epsilon$,%
\begin{eqnarray}
\Omega_{\mathrm{Ch}} & = &
\,\Omega_{\mathrm{Ch}0}\left[\frac{\bar{\Omega}_{\Lambda0}}{\Omega
_{\mathrm{Ch}0}}+\frac{\bar{\Omega}_{m0}}{\Omega_{\mathrm{Ch}0}}(1+x)^{3\epsilon}\right]
^{1/\epsilon},\nonumber\\
& = &\,\Omega_{\mathrm{Ch}0}\left[  1+\frac{\bar{\Omega}_{m0}}{\Omega_{\mathrm{Ch}0}}%
3\log(1+x)\epsilon\right]  ^{1/\epsilon}.
\end{eqnarray}
We can now take the limit $\epsilon\rightarrow0$ in the above
equation, substitute in eq.~(\ref{chiGCG2}), integrate, and solve
for $\Omega_{\mathrm{Ch}0}$ to finally obtain

\begin{eqnarray}
\label{condm1}%
\bar{\Omega}_{m0}+\bar{\Omega}_{\Lambda 0} & = &
\sec^{2}(K\,r_{inj})  +\tan^{2} (K\, r_{inj})
\,\bar{\Omega}_{\Lambda0}\;,\nonumber\\ &&\quad\quad\quad\quad
\mathrm{for}\quad\Omega_{\mathrm{Ch}0}<1\;\mathrm{and}\;K>0,\nonumber\\
\bar{\Omega}_{m0}+\bar{\Omega}_{\Lambda 0} & = &
\mathrm{sech}^{2}( K\,r_{inj}) -\tanh ^{2}(K\,r_{inj})
\,\bar{\Omega}_{\Lambda0}\;,\\ &&\quad\quad\quad\quad
\mathrm{for}\quad\Omega_{\mathrm{Ch}0}>1\;\mathrm{and}\;K>0,\nonumber\\
\bar{\Omega}_{m0}+\bar{\Omega}_{\Lambda 0} & = &
1\;,\quad\quad\;\;\,\mathrm{for}\quad K\leq0,\nonumber
\\&   &
\quad\quad\quad\quad\quad\quad\quad\quad\quad\quad\quad\quad
(\alpha=-1)\nonumber
\end{eqnarray}
where $K=\left(3\bar{\Omega}_{m0}/2\Omega_{\mathrm{Ch}0}-1\right)$.
Here again, as well as in eq.~(\ref{cond0}) below, the
undetectability conditions can be derived from the contour curves by
replacing the equalities by inequalities, as in eq.~(\ref{condinf2}).
Note that for current observational values
$K\leq0$, in which case \emph{any} non-flat topology would be
detectable in principle. This may seem surprising, but is a
consequence of the fact that in this range the contour curve and the
flat line $\Omega_{\mathrm{Ch}0}=1$ coincide, and thus there is no
undetectable region. As we shall show, however, even a small
non-zero value of $\Omega_{b0}$ renders many topologies
unobservable.

Finally, as shown by Mota et al. (\cite{Mota03}), the contour curve
for the $\alpha=0$ (or $\Lambda$CDM) case can be well approximated
by the secant line, which is given by

\begin{eqnarray}
\label{cond0}%
\bar{\Omega}_{m0}+\bar{\Omega}_{\Lambda 0} & = &
\mathrm{sech}^{2}\left(
\frac{r_{inj}}{2}\right)+\tanh^{2}\left(\frac{r_{inj}}{2}\right)
\bar{\Omega}_{\Lambda0}\nonumber\\ &&\quad\quad\quad\quad
\mathrm{for}\;\; \Omega_{\mathrm{Ch}0}<1,\nonumber\\
\bar{\Omega}_{m0}+\bar{\Omega}_{\Lambda 0} & =&
\mathrm{sec}^{2}\left(
\frac{r_{inj}}{2}\right)-\tan^{2}\left(\frac{r_{inj}}{2}\right)
\bar{\Omega}_{\Lambda0}\quad\\ &&\quad\quad\quad\quad
\mathrm{for}\;\; \Omega_{\mathrm{Ch}0}>1.\nonumber\\&   &
\quad\quad\quad\quad\quad\quad\quad\quad\quad\quad\quad\quad\quad
(\alpha=0)\nonumber
\end{eqnarray}

\begin{figure}
   \centering
   \includegraphics[width=8.5cm]{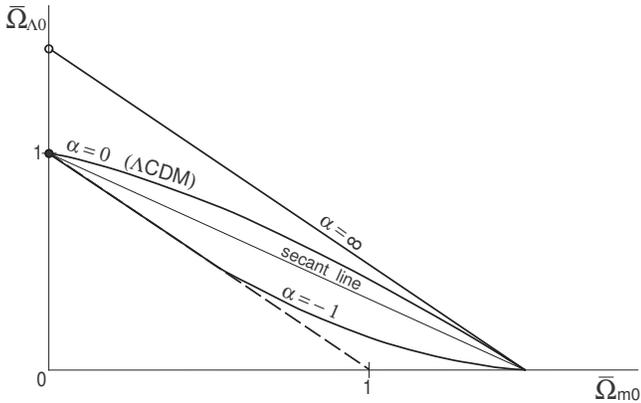}
      \caption{Schematic contour curves $\chi_{obs}=r_{inj}^{M}$ for
      $\alpha=-1,0$ and $\infty$, with $z_{obs}=\infty$
       and $\Omega_{b0}=0$. The curves for intermediate values of
       $\alpha$ will interpolate monotonically between these extremes.
       Using $z_{obs}=1089$ instead does not change the curves
       significantly. The presence of baryonic matter prevents the
       $\alpha=-1$ curve from touching the flat line (dashed). The
       secant line (thin line) is a good approximation of the
       $\alpha=0$ curve.
              }
         \label{curves}
\end{figure}

Figure \ref{curves} illustrates schematically the detectability
using contour curves in the parametric plane
$\bar{\Omega}_{m0}-\bar{\Omega}_{\Lambda0}$. We portray the
contour curves for a spherical manifold associated with
$\alpha=-1,0$ and $\infty$, as given above, as well as the secant
line. For each of these cases, if $\bar{\Omega}_{m0}$ and
$\bar{\Omega}_{\Lambda0}$ take values between the respective
contour curve and the flat line (dashed), then the topology of $M$
is undetectable.

For intermediate values of $\alpha$ we must turn to numerical
integration to compute $\chi_{obs}$ and study quantitatively the
effects of the GCG parameters on the detectability of the topology.
We consider the set of topologies presented in Tables 1 and 2, and
assess their (un)detectability based on current observational values
of the cosmological density parameters and bounds on the parameter
$\alpha$ of the GCG. We are particularly interested in the effect of
the value of $\alpha$ on the potential detectability of the
topology, as different values of $\alpha$ can be thought of as
different models for the matter-energy content.

In the numerical computations we have set $\Omega_{b0}=0.04$ in equation
(\ref{chiGCG2}). This value can be obtained from observations of
light element abundances, combined with primordial Big-Bang
nucleosynthesis analysis (Burles et al. \cite{BBN}, Kirkman et al.
\cite{DH}) and the value of the Hubble constant (Freedman
\cite{freedman}). For $\Omega_{0}$ we use the limits from a
combination of CMBR and large-scale structure data obtained by Tegmark et al.
(\cite{SDSSWMAP}) for the $\Lambda$CDM case, $0.99<\Omega
_{0}<1.03$. Finally
we fix the total matter-like density parameter at $\bar{\Omega}_{m0}%
+\Omega_{b0}=0.3$. Our results however are not very sensitive to the
precise value of $\bar{\Omega}_{m0}$. Indeed, for values of $\alpha$
not too negative ($\alpha>-0.5$), we have
$\Delta\chi_{obs}/\chi_{obs}\simeq 0.15\,%
\Delta\bar{\Omega}_{m0}/\bar{\Omega}_{m0}$.

\begin{figure}
   \centering
   \includegraphics[width=9.5cm]{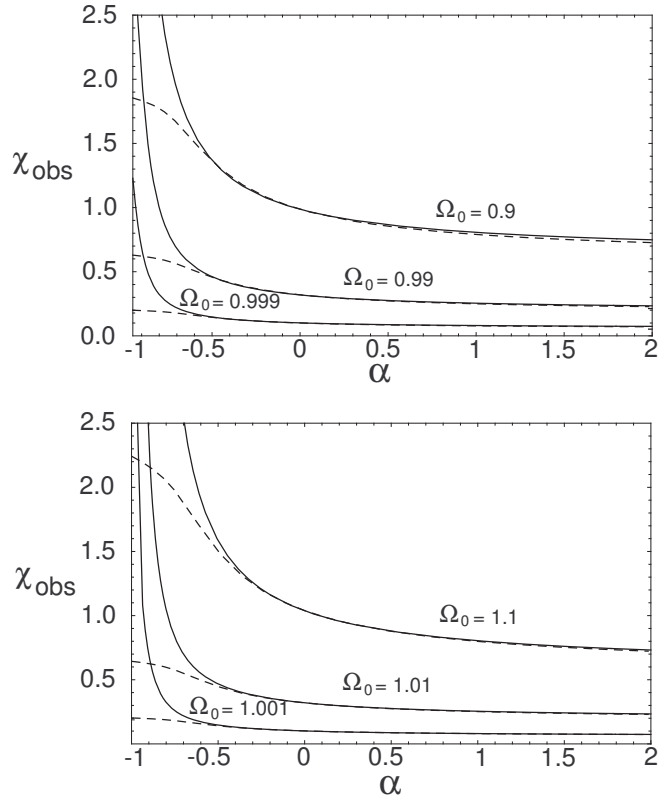}
      \caption{Survey depth $\chi_{obs}$, for several values of
                $\Omega_{0}=\Omega_{\mathrm{Ch}0}+\Omega_{b0}$, for hyperbolic (top) and spherical (bottom)
                universes, with $z_{obs}=1089$, and
                $\bar{\Omega}_{m0}+\Omega_{b0}=0.3$. The baryonic term is
                neglected in the solid lines and taken into account in the
                dashed lines.
              }
         \label{chixalpha}
\end{figure}

To quantify the dependency on $\alpha$ and $\Omega_{b0}$, we plot in
figure (\ref{chixalpha}) the depth $\chi_{obs}$ as a
function of $\alpha$, for various values of $\Omega_{0}$. The plot
with $\Omega_{b0}=0$ is shown in solid lines, and the plot with
$\Omega_{b0}=0.04$ is given by the dashed lines.

The plots make clear a number of things. First, the value of
$\chi_{obs}$ is not very sensitive to $\alpha$ for $\alpha>0$.
Moreover, the change is very small for $\alpha>1$
($\Delta\chi_{obs}/\chi_{obs}\simeq0.3$ in the range
$1\leq\alpha\leq\infty$). This implies that the undetectability
conditions %
(\ref{condinf2}), which were derived for $\alpha \rightarrow \infty$,
are not much more restrictive than the
undetectabilty condition obtained in the case of the standard
Chaplygin gas (i.e., $\alpha=1$). Note also that the presence of a
small $\Omega_{b0}$ component does not change this picture in a
significant way for positive values of $\alpha$. On the other hand,
$\chi_{obs}$ changes substantially when $\alpha<0$. This was
expected, since we have shown that if $\Omega_{b0}=0$ then
$\chi_{obs}$ diverges in the limit $z\rightarrow\infty$ when
$\alpha=-1$. The presence of a small amount of baryonic matter
prevents this divergence, although $\chi _{obs}$ still changes
appreciably for $\alpha<0$. It is clear then that for sufficiently
negative values of $\alpha$, the precise value of $\Omega_{b0}$
plays an important role in the detection of cosmic topology in the
GCG context.

In table \ref{tabtop} we display the manifolds from
tables~\ref{SingleAction} and~\ref{10HW-Census}, along with their
respective injectivity radii. The upper (for spherical manifolds)
and lower (for hyperbolic ones) bounds on $\Omega_{0}$ so that the
manifold's topology is detectable in principle are shown in the
table, for some noteworthy values of $\alpha$. We take the
theoretical lower bound $\alpha  = -1$; a numerical lower bound
$\alpha  = -1/2$ obtained from combining several observables and
assuming a flat universe (see Makler et al. \cite{GCG2} and Zhu
\cite{Zhu}); the $\Lambda$CDM case $\alpha = 0$; and the standard
Chaplygin Gas $\alpha  = 1$. The manifolds which are undetectable
for $0.99\leq\Omega_{0}\leq1.03$ are indicated with boldface.
%
\begin{table}[!htb]
\caption{Minimum (maximum) values of $\Omega_{0}$ for each
spherical (hyperbolic) topology to be potentially detectable (for
$\bar{\Omega}_{M}=0.26,$ $\Omega_{b}=0.04$, and $z=1089$). Numbers
in \textbf{boldface} indicate undetectabilty for the corresponding
topologies.}\label{tabtop}
\centering
\begin{tabular}{ l c c c c c c}
\hline\hline {} &{} &{} &{} &{} &{}\\
[-1.5ex] Manifolds & $r_{inj}$ & $\alpha=-1$ & $\alpha=-0.5$ &
$\alpha=0$ & $\alpha=1$ \\
[1ex] \hline {} &{} &{} &{} &{} &{}\\[-1.5ex]
$D_{9}^{\ast},\ Z_{18}$ & $\frac{\pi}{18}$ & 1.001 & 1.002 & 1.003
& 1.004\\
[1ex] $D_{6}^{\ast},\ Z_{12}$ & $\frac{\pi}{12}
$& 1.002 & 1.004 & 1.007 & 1.010\\
[1ex] $I^{\ast},D_{5}^{\ast},\ Z_{10}$ & $\frac{\pi}{10}$ & 1.003
& 1.005 & 1.010 & 1.015\\
[1ex] $O^{\ast},D_{4}^{\ast},\ Z_{8}$ & $\frac{\pi}{8}$ & 1.004 &
1.008 & 1.015 & 1.023\\
[1ex] $T^{\ast},D_{3}^{\ast},\ Z_{6} $&$ \frac{\pi}{6}$ &
1.007 & 1.014 & 1.026 & \textbf{1.042}\\
[1ex] $D_{2}^{\ast},\ Z_{4}$ & $\frac{\pi}{4}$ & 1.016 &
\textbf{1.031} & \textbf{1.059} & \textbf{1.094}\\
[1ex]\hline {} &{} &{} &{} &{} &{}\\
[-1.5ex] m004(1,2) & 0.183 & 0.999 & 0.998 & 0.997 & 0.995\\
[1ex] m004(6,1) & 0.240 & 0.998 & 0.997 & 0.994 & 0.991\\
[1ex] m003(-4,3) & 0.288 & 0.998 & 0.996 & 0.992 & \textbf{0.987}\\
[1ex] m003(-2,3) & 0.289 & 0.998 & 0.996 & 0.992 & \textbf{0.987}\\
[1ex] m003(-3,1) & 0.292 & 0.998 & 0.996 & 0.992 & \textbf{0.987}\\
[1ex] m009(4,1) & 0.387 & 0.996 & 0.992 & \textbf{0.985}
& \textbf{0.977}\\
[1ex] m007(3,1) & 0.416 & 0.995 & 0.991 & \textbf{0.983}
& \textbf{0.974}\\
[1ex] \hline
\end{tabular}
\end{table}

The table shows that the choice of $\alpha$ plays an important role
in deciding on the detectability of these manifolds. In particular,
negative values of $\alpha$ greatly favor detectability. All the
manifolds above are potentially detectable for $\alpha=-1$, and all
but $D_{2}^{\ast}$ and $Z_{4}$ are for $\alpha=-1/2$. In the case
of $\alpha=1$, most of the selected hyperbolic and many of the singe
action spherical manifolds are undetectable. It is clear that the
detectability of individual topologies depends significantly on the
value of $\alpha$, and in particular may greatly differ from the
case of a $\Lambda$CDM model. 
This is an important point, inasmuch as it makes explicit the
dependence of the detectability with respect to the model for the
matter-energy content.

Of course one cannot list all the (infinite) manifolds in the
dihedral and cyclic groups. These two classes are particularly
important because as $\Omega_{0}\rightarrow1$ from above, there is
always an $n_{\ast}$ (or $m_{\ast}$) such that the topology
corresponding to $Z_{n}$ (or $D_{m}^{\ast}$) is detectable in
principle for $n>n_{\ast}$ (or $m>m_{\ast}$). We have seen that
(\ref{condinf}) provides a lower bound for $\chi_{obs}$ as a
function of $\alpha$ and is also a fair approximation for the
contour curves when $\alpha\geq1$ for these manifolds. Solving
(\ref{condinf}) for $n_{\ast}$ and $m_{\ast}$ we can state that in
a dihedral or lens space, for any value of $\alpha$ the topology
is detectable (in principle) if
\begin{eqnarray}
\label{absnm}%
n_{\ast} &  \geq &\mathrm{int}\left[ \frac{\pi}{2}\frac{1}{\arctan
(\sqrt{\Omega_{\mathrm{Ch}0}-1})}\right]\nonumber,
\\ m_{\ast} &  \geq &\mathrm{int}\left[ \pi\frac{1}{\arctan(\sqrt{\Omega
_{\mathrm{Ch}0}-1})}\right].
\end{eqnarray}
More specifically, for $\alpha\geq 1$ these are approximately
equalities, since, as can be seen in Fig. \ref{chixalpha},
$\chi_{obs}$ is not very sensitive to the value of $\alpha$ for
$\alpha\geq1$. Recall that, although these expressions do not take
baryons into account, their contribution is negligible for
$\alpha\geq 1$. Thus, eqs.~(\ref{absnm}) provide an absolute lower
bound for the detectability of cyclic and dihedral topologies. Any
values of $n$ and $m$ greater than $n_{\ast}$ and $m_{\ast}$,
respectively, will render the topology potentially detectable for
any value of $\alpha$.

Given the increasing amount of high quality cosmological data, in
particular the availability of high resolution full sky CMBR maps
(see Spergel et al. \cite{WMAP-Spergel-et-al}), we can expect to
detect a non-trivial topology if it is present and if
$r_{inj}<\chi_{obs}$. This can be done, for example, by observing
matching circles in the CMBR maps.
Indeed some topological signatures have recently
been proposed and tested (see Luminet et al. \cite{Poincare},
Cornish et al. \cite{Cornishetal03}, Roukema et al. \cite{Roukema},
and Aurich et al \cite{Aurich}). One can then ask whether the
hypothetical detection of a nontrivial topology may lead to a better
knowledge of the cosmological model. In the case of the GCG, does
the determination of a given topology impose any constraint on
$\alpha$?

\begin{figure}
   \centering
   \includegraphics[width=8.6cm]{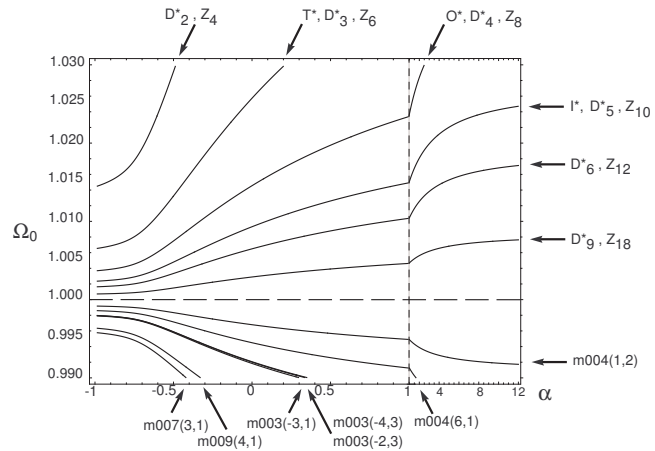}
      \caption{Curves of constant $\chi_{obs}=r_{inj}^{\;M}$, in the
                ($\Omega_{0}$, $\alpha$) plane for the same topologies discussed
                in the previous section. The $\alpha$ axis is compressed for
                $1\leq\alpha\leq12$. Again we fix $z=1089$, and
                $\bar{\Omega}_{m0}+\Omega _{b0}=0.3$, with $\Omega_{b0}=0.04$.
              }
         \label{Omal}
\end{figure}

The answer is positive, as can be seen from figure (\ref{Omal}).
There we plot the contour curves $\chi_{obs}(\Omega_0,\alpha)
=r_{inj}^{\;M}$, in the $\Omega_{0}- \alpha$ parametric plane for
the same topologies discussed in the previous section, where we
fix $z_{obs}=1089$, $\Omega_{b0}=0.04$ and
$\bar{\Omega}_{m0}=0.26$. Recall from section (\ref{detect}) that
each such contour curve separates the parametric plane in two
regions where the manifold in question is respectively potentially
detectable and undetectable. It is clear, for example, that the
detection of a binary tetrahedral topology ($T^{\ast}$) would
place the constraint $\alpha<-0.2$ for current bounds on
$\Omega_{0}$. If future observations tighten the range of
$\Omega_{0}$ even closer to $1$, let's say
$\Omega_{0}=1.000\pm0.003$, the detection of either
$D_{9}^{\ast},\ Z_{18}$, m$004(1,2)$, or m$004(6,1)$ would imply
$\alpha\lesssim-0.5$.
Figure (\ref{Omal}) shows that a topology that would be detectable
in a $\Lambda$CDM universe is not necessarily ruled out if no
observational topological signature is found. For example, if no
sign is detected of the binary octahedral ($O^{\ast}$) topology in
CMBR maps, it could simply be that $\alpha\gtrsim 2.5,$ so that
this topology is in fact unobservable in the $\Lambda$CDM context.

\section{Discussion and concluding remarks\label{conclusion}}

We have considered the detectability of a non-trivial cosmic
topology in a universe dominated by the GCG. This component offers
the possibility of dark-matter/energy unification and is consistent
with a number of cosmological observations. We have investigated how
the detectability is altered with respect to changes in the model
parameters, as well as to the choice of the model itself, using the
GCG family as a concrete example.

In the case of the $\Lambda$CDM model, the detectablity of a
manifold's topology is a function of the component densities,
$\Omega_{m0}$ and $\Omega_{\Lambda 0}$. The consideration of the
GCG brings in a new parameter into the analysis, the steepness of
the equation of state $\alpha$. We have shown that, for fixed
$\Omega_{0}$ and $\bar{\Omega}_{m0}$, more topologies become
potentially detectable as $\alpha$ decreases. We have obtained
analytical results for $\alpha\rightarrow\infty$, establishing a
lower bound for the detectability in the GCG case. In general, the
detectability is not very sensitive to $\alpha$, for $\alpha>0$,
but varies greatly for negative values of $\alpha$. In this latter
case, the contribution of baryons is important and must be taken
into account. For example, we have shown that for $\alpha=-1$ any
topology would be potentially detectable if we neglect
$\Omega_{b0}$. However, considering the baryonic contribution,
many manifolds would still be unobservable.

It is now possible to expect detectable non-trivial topologies to
leave a measurable imprint on CMBR maps. The detection of a cosmic
topology is therefore a realistic possibility in the near future,
and we have investigated what constraints would be imposed on the
parameters of the model, by such a detection. For example, the
hypothetical detection of a binary tetrahedral topology would place
the constraint $\alpha<-0.2$ for $\Omega_{0}<1.02$,
$\bar{\Omega}_{m0}=0.26$, and $\Omega_{b}=0.04$. A more thorough
study of the inverse problem is developed in  Makler et al.
(\cite{paperGCGtopology}).

In the present work we outline a method for studying the
detectability of cosmic topology systematically for a family of
cosmological models. Although this method was applied to the GCG
model for DM and DE unification, it can be used to study other
models or choices of variables. For instance,
a more physically motivated parametrization might be to use $\Omega_{m0}^{eff}%
=\Omega_{\mathrm{Ch}0}\left(  1-A\right)  ^{1/\left( 1+\alpha\right)
}+\Omega_{b}$, which is the fraction that behaves as CDM for early
times and is measured from the matter clustering. This variable is
more directly constrained by some observational data (see Makler et
al. \cite{GCG2} and Reis et al. \cite{entropy}), and may be better
suited for studies involving comparisons with observational
constraints. This choice of variables is adopted in Makler et al.
(\cite{paperGCGtopology}).

An important point that emerges from these results is that given a
set of observational constraints, the detectability of cosmic
topology depends on the choice of the cosmological
model. Thus, any attempt to rule out a given topology must take
this into account.

\begin{acknowledgements}
    We thank MCT, CNPq, and FAPERJ for the grants under which this work was
    carried out.
\end{acknowledgements}

\end{document}